\documentclass[11pt,a4paper]{article}%

\usepackage{amsmath,amsfonts,amsbsy}

\usepackage{dsfont}

\usepackage{simplewick}
\usepackage{array}
\usepackage{color}
\usepackage{graphicx}
\usepackage[vcentermath]{youngtab}
\usepackage[all]{xy}
\usepackage{graphicx,epsfig}
\usepackage{amsfonts}
\usepackage{amssymb}
\usepackage{multirow}
\usepackage{caption}
\usepackage{graphics}
\usepackage{graphicx}
\usepackage{float}
\usepackage{verbatim}

\usepackage{amsmath,amssymb,amsfonts}
\usepackage{graphicx,epsfig}
\usepackage{color}
\usepackage{array}
\usepackage{amsmath}
\usepackage{booktabs}
\usepackage{amsfonts}
\usepackage{amssymb}
\usepackage{multirow}
\usepackage{float}
\usepackage{cancel}
\usepackage{caption}
\usepackage{graphics}
\usepackage{graphicx}
\usepackage{jheppub}

\def\be{\begin{equation}}
\def\ee{\end{equation}}
\def\ba{\begin{eqnarray}}
\def\ea{\end{eqnarray}}

\def\m{\mu}
\def\D{\Delta}
\def\n{\nu}

\def\p{\partial}
\def\nn{\nonumber}
\begin{document}

\begin{titlepage}
\begin{flushright}
\end{flushright}
\vskip 1.0cm
\begin{center}
{\Large \bf Improved bounds for CFT's with global symmetries} \vskip 1.0cm
{\large \ Alessandro Vichi$^a$} \\[1cm]
{\it $^a$ Institut de Th\'eorie des Ph\'enom\`enes Physiques, EPFL,  CH--1015 Lausanne, Switzerland}\\[5mm]
\vskip 1.0cm 
{\bf Abstract}
\end{center}
\begin{abstract}

The four point function of Conformal Field Theories (CFT's) with global symmetry gives rise to multiple crossing symmetry constraints. We explicitly study the correlator of four scalar operators transforming in the fundamental representation of a global $SO(N)$ and the correlator of chiral and anti-chiral superfields in a superconformal field theory. In both cases the constraints take the form of a triple sum rule, whose feasibility can be translated  into restrictions on the CFT spectrum and interactions. In the case of $SO(N)$ global symmetry we derive bounds for the first scalar singlet operator entering the Operator Product Expansion (OPE) of two fundamental representations for different value of $N$. Bounds for the first scalar traceless-symmetric representation of the global symmetry are computed as well. Results for superconformal field theories improve previous investigations due to the use of the full set of constraints.  Our analysis only assumes unitarity of the CFT, crossing symmetry of the four point function and existence of an OPE for scalars. 
\end{abstract}
\end{titlepage}

\newpage



\section{Introduction}
\label{Introduction}

Conformal Field Theories are extensively used in condensed matter and in String theory. The remarkable achievements in those fields   have been obtained due to the special feature of conformal algebra in two dimensions, which can be extended to the so called Virasoro algebra (\cite{Belavin:1984vu}). This infinite dimensional algebra provides a huge amount of additional information, allowing for instance to completely solve for specific models (\cite{Belavin:1984vu}).

Exact results for CFT's in dimensions larger then two are much harder to get because of the lack of an higher dimensional version of the Virasoro algebra. Until few years ago only few examples of CFT's were known, either based on perturbative results, such as  \cite{BZ,BM} or on supersymmetry (\cite{Seiberg:1994pq,Argyres:1995jj}). Generically, however, it is not known how to extract the spectrum of the theory or how to compute all the correlation functions. Nevertheless it has been conjectured (\cite{pol}) that imposing the right consistency conditions on the CFT is sufficient to solve it completely, exactly as in two dimensions. This takes the name of the \emph{bootstrap program}. Unfortunately, despite the significant amount of works on the subject, very few exact results have been achieved in this direction. 

A new technique to investigate CFT's has been introduced in \cite{r1}, where a general procedure to extract informations about CFT's is discussed. The authors of \cite{r1} made use of the works of Dolan \& Osborn \cite{Dolan:2003hv}, to translated the  obscure crossing symmetry constraints into precise functional equations, taking the form of functional sum rules. Studying the feasibility of these constraints it is possible to obtain informations about the spectrum and the size of OPE coefficients of a general CFT. The first two works \cite{r1,r2} focused on CFT's containing at least a real scalar field of dimension $d$. Under this condition, assuming only unitarity, they showed that the CFT must contain a scalar operator with dimension smaller than a certain function depending on $d$. The possibility to extract upper bounds on OPE coefficients has been shown in \cite{cr}. Interestingly, some OPE coefficients can be related to central charges: in \cite{r3,poland} this techniques has been used to derive lower bounds on those quantities. \cite{poland} also generalized the method to superconformal field theories containing at least a chiral scalar superfield. All the results extracted so far are based on a single crossing symmetry constraint. However, as shown in \cite{r4}, when the theory posses global symmetries, it is possible to derive several crossing symmetry relation, namely multiple  sum rules. Because of a complication in the numerical algorithm exploited for the bound computation, we were not able to produce numerically relevant results in \cite{r4}. In the present work we present bounds for CFT's with $SO(N)$ global symmetry and for superconformal field theories making use of the full set of constraints. In both these class of theories the four point functions gives rise to a triple sum rule. In the next section we review the crossing symmetry constraints for the case of a single real scalar and for $SO(N)$ theories. In Section \ref{Superconformal field theories} we discuss superconformal field theories, with particular attention to the operators allowed to enter the OPE's of chiral superfields, and we derive explicitly the triple sum rule. We present numerical results in Section \ref{Bounds and numerical results}. In the context of $SO(N)$ theories we will show that it is possible to derive independent bounds on the dimensions of scalar operators belonging to different representations of the symmetry group. Concerning SCFT, we derive bounds on the first non-chiral operator entering the OPE of a chiral superfield and we compute a lower bound on the central charge. Compared with the results of \cite{poland}, our results appear stronger since we are exploiting all the equations.

All the results presented in this work has been produced using the procedure explained in the appendix of \cite{r2}. This in particular requires to reduce to a discrete Linear Programming problem. We will discuss in the conclusions possible alternatives to this method.

\section{The bootstrap equation}
\label{The bootstrap equation}

We begin with some preliminary comments and notational conventions. We will
work in the $D=4$ Euclidean space.	\\
Consider a 4-point function containing four scalar primary operators, not necessarily Hermitian, assumed to have equal dimensions $d$:
$\left\langle \phi_1(x_{1})\phi_2(x_{2}) \phi_3(x_{3})\phi_4(x_{4})\right\rangle $. 
The OPE of two of those fields $\phi_1\times \phi_2$ will contain a sequence of spin $l,$
dimension $\Delta$ primary fields $O_{\Delta,l}$,:
\begin{equation}
\phi_i\times\phi_j=\sum_{\Delta,l} c^{ij}_{\Delta,l}O_{\Delta,l}\,.
\end{equation}
and of course all their descendants. The OPE coefficients  $c^{ij}_{\Delta,l}$ and the operators $O_{\Delta,l}$ depends on the choice of $i$ and $j$.
Here $c^{ij}_{\Delta,l}$, in general complex, are meant as the normalization of the three point function $\langle \phi_i\phi_j O_{\Delta,l}\rangle$. As shown in \cite{Dolan:2003hv}, the four point function can be written as a sum of conformal blocks, each of them encoding the contribution to the four point function of a primary operator and all its descendants.  Depending on how the OPE is taken inside the correlation function we get different decomposition. For instance, the (12)(34) channel ($\equiv$s-channel) decomposition gives
\begin{gather}
\langle
	 \bcontraction[1ex]{}{\phi_1(x_1)}{}{\phi_2(x_2)}
	 \bcontraction[1ex]{\phi_1(x_1)\phi_2(x_2)}{\phi_3(x_3)}{}{\phi_4(x_4)}  
	 \phi_1(x_1)\phi_2(x_2)\phi_3(x_3)\phi_4(x_4) 
	 \rangle=\sum_{\Delta,l}\frac{1}{x_{12}^{2d}x_{34}^{2d}}\,p_{\Delta
,l}\,g_{\Delta,l}(u,v)\,,\label{eq:confblocks}\\
u\equiv x_{12}^{2}x_{34}^{2}/(x_{13}^{2}x_{24}^{2})=z\bar{z},\quad v\equiv
x_{14}^{2}x_{23}^{2}/(x_{13}^{2}x_{24}^{2})\,=(1-z)(1-\bar{z})\,,\\
g_{\Delta,l}(u,v)=+\frac{z\bar{z}}{z-\bar{z}}[k_{\Delta+l}(z)k_{\Delta
-l-2}(\bar{z})-(z\leftrightarrow\bar{z})]\,,\\
k_{\beta}(x)\equiv x^{\beta/2}{}_{2}F_{1}\left(  \beta/2,\beta/2,\beta
;x\right)  \,.
\end{gather}
The above equations fixes our conventions for the conformal blocks. The coefficients $p_{\Delta,l}$ are given by
\begin{equation}
p_{\Delta,l}=\frac{c^{12}_{\Delta,l}c^{34}_{\Delta,l}}{2^l}%
\end{equation}
Compared to \cite{Dolan:2003hv}, and also to \cite{r1,r2}, we have dropped the
$(-1/2)^{l}$ pre-factor in the expression for $g_{\Delta,l}$. 
Notice that if we choose $\phi_3$ and $\phi_4$ to be the Hermitian conjugate of $\phi_1$ and $\phi_2$ the above coefficients become definite positive.

The conformal block decomposition in the (14)(23) channel ($\equiv$t-channel) must produce the same result. As shown in \cite{r1,r2,r3,r4,cr,poland}, equating different channel gives rise to non trivial crossing symmetry constraint. We will briefly review the form of this constraint for the case of a four point function containing four equal real scalars transforming in a fundamental representation of a global $SO(N)$ symmetry. Then in Sec. \ref{Superconformal field theories} we will discuss the case of complex scalar field in a superconformal field theory.

\subsection{The simplest case: Sum rule without symmetries}
\label{The simplest case: Sum rule without symmetries}

We start the analysis focusing on the particular case when $\phi_i$ is Hermitian,
$\phi_i=\phi$  and $d=[\phi]$(\cite{r1,r2}).  In this case the s- and t-channels of the four point function:
\ba\label{eq:dec}
	&&\langle \bcontraction[1ex]{}{\phi(x_1)}{}{\phi(x_2)}
	 \bcontraction[1ex]{\phi(x_1)\phi(x_2)}{\phi(x_3)}{}{\phi(x_4)}  
	 \phi(x_1)\phi(x_2)\phi(x_3)\phi(x_4) 
	 \rangle=\frac{1}{x_{12}^{2d}x_{34}^{2d}}\left(1+\sum_{\Delta,l}\,p_{\Delta
,l}\,g_{\Delta,l}(u,v)\right),\\
	&&\langle \bcontraction[1ex]{}{\phi(x_1)}{\phi(x_2)\phi(x_3)}{\phi(x_4)}
	 \bcontraction[2ex]{\phi(x_1)}{\phi(x_2)}{}{\phi(x_3)}  
	 \phi(x_1)\phi(x_2)\phi(x_3)\phi(x_4) 
	 \rangle=\frac{1}{x_{14}^{2d}x_{23}^{2d}}\left(1+\sum_{\Delta,l}\,p_{\Delta
,l}\,g_{\Delta,l}(v,u)\right)\,,
\ea
correspond to the same OPE
($\phi\times\phi$). The only change is the pre-factor and the $u,v$ dependence of the conformal blocks
(other permutations do not give additional information \cite{r1,r2}). 
Notice that we explicitly separated the contribution of the unit
operator, present in the $\phi\times\phi$ OPE. We stress that all conformal blocks appear in
(\ref{eq:dec}) with positive coefficients.\\
All the operators appearing in the OPE $\phi\times\phi$ have even
spin\footnote{A formal proof of this fact can be given by considering the
3-point function $\left\langle \phi(x)\phi(-x)\mathcal{O}_{(\mu)}%
(0)\right\rangle$. See \cite{r1} for more details.}. The constraint imposed by crossing symmetry 
can be expressed \cite{r1} in the form of the following \textit{sum
rule}:%
\ba\label{eq:F}
& 1=\sum_{\Delta,l}p_{\Delta,l}F_{d,\Delta,l}(z,\bar{z}),\quad p_{\Delta ,l}>0\,, \nn\\
& F_{d,\Delta,l}(z,\bar{z})\equiv\frac{v^{d}g_{\Delta,l}(u,v)-u^{d}g_{\Delta,l}(v,u)\,}{u^{d}-v^{d}},
\label{sumrule}%
\ea
where the sum is taken over all $\Delta,l$ corresponding to the operators
$\mathcal{O}\in\phi\times\phi$.\\
To get an idea about what one can expect from the sum rule, we can consider the free scalar theory. In this case $d=1$,
and only operators of twist $\Delta-l=2$ are present in the OPE $\phi
\times\phi$,\cite{Dolan:2003hv}.The OPE coefficients of all these operators (or rather their squares) can be
found by decomposing the free scalar 4-point function
 into the corresponding conformal blocks. This gives \cite{Dolan:2003hv}
\begin{equation}
c_{l}=(1+(-)^{l})\frac{(l!)^{2}}{(2l)!} \label{eq:noglobal-coeffs}%
\end{equation}

\subsection{CFT's with Global symmetries}
\label{CFT's with Global symmetries}

We now discuss a generalization of the sum rule introduced in the previous section to
the case when the CFT has a continuous global symmetry $SO(N)$ and the operator $\phi_a$ transforms in the fundamental representation.
It is useful to recall that the original motivation of \cite{r1} was to find a
bound of precisely this type for the case of $SO(4)$. This in turn was needed in order to constrain the Conformal
Technicolor scenario of electroweak symmetry breaking \cite{luty}. 

We normalize the 2-point function of $\phi_{a}$ as
$\left\langle \phi_{a}(x)\phi_{b}(0)\right\rangle =\delta_{ab}\left(
x^{2}\right)  ^{-d}$, $d=[\phi]$. Consider the 4-point function $\left\langle \phi_a(x_{1})\phi_b(x_{2}) \phi_c(x_{3})\phi_d(x_{4})\right\rangle $
Operators appearing in the $\phi_{a}\times\phi_{b}$ OPE can transform under
the global symmetry as singlets $S$, symmetric traceless tensors $T_{(ab)}$,
or antisymmetric tensors $A_{[ab]}$:
\begin{align}
\phi_{a}\times\phi_{b}  &  =\delta_{ab}\mathds{1}  +\delta_{ab}S^{(\alpha)}\quad\text{(even spins)} +T_{(ab)}^{(\alpha)}\quad\text{(even spins)} +A_{[ab]}^{(\alpha)}\quad\text{(odd spins)}\label{OPEson}
\end{align}
The index $\left(  \alpha\right)  $ shows that an arbitrary number of
operators of each type may in general be present, of various dimensions
$\Delta$ and spins $l$. However, permutation symmetry of the $\phi_{a}\phi
_{b}$ state implies that the spins of the $S$'s and $T$'s will be even, while
they will be odd for the $A$'s. It will be important for us that the unit operator
$\mathds{1}$\ is always present in the $\phi_{a}\times\phi_{b}$ OPE, with a
unit coefficient.\\
Equating the conformal block decomposition in the s-channel and the t-channel and matching the different index structures we can obtain three non trivial crossing symmetry relations \cite{r3} which can conveniently rewritten in the form of a vectorial sum rule:
\be\label{eq:son-vect}
\sum p_{\Delta,l}^{S}
\underbrace{
\left(
\begin{array}
[c]{c}%
0\\
F_{d,\Delta,l}\\
H_{d,\Delta,l}%
\end{array}
\right)}_{\vec V^S_{\Delta,l}}
 +\sum p_{\Delta,l}^{T}
 \underbrace{
 \left(
\begin{array}
[c]{c}%
F_{d,\Delta,l}\\
\left(  1-\frac{2}{N}\right)  F_{d,\Delta,l}\\
-\left(  1+\frac{2}{N}\right)  H_{d,\Delta,l}%
\end{array}
\right)}_{\vec V^T_{\Delta,l}}
  +\sum p_{\Delta,l}^{A}
 \underbrace{ 
  \left(
\begin{array}
[c]{c}%
-F_{d,\Delta,l}\\
F_{d,\Delta,l}\\
-H_{d,\Delta,l}%
\end{array}
\right) }_{\vec V^A_{\Delta,l}}
 =
\underbrace{
 \left(
\begin{array}
[c]{c}%
0\\
1\\
-1
\end{array}
\right)}_{\vec V^{RHS}}
\ee %
where we have introduced the function
\be\label{eq:H}
\  H_{d,\Delta,l}(z,\bar{z})\equiv\frac{v^{d}g_{\Delta,l}(u,v)+u^{d}g_{\Delta,l}(v,u)\,}{u^{d}+v^{d}},
\ee
The quantities $p_{\Delta,l}^{S,T,A}$ are related to the OPE coefficient of operator transforming in one of the $SO(N)$ representations (\ref{OPEson}).\\
Again it is interesting to find an $SO(N)$ decomposition of the 4-point function in
the theory of $N$ free real scalars. The decomposition of the four point function in conformal block this time gives \cite{r4}
\ba
&& p^{T},p^{A}=\frac{(l!)^{2}}{(2l)!}\text{ (}l\text{ even/odd)}\,,\qquad  p^{S}=\frac{2}{N}\frac{(l!)^{2}}{(2l)!}.
\ea
One can check what with these coefficients the triple sum rule converges
rapidly near $z=\bar{z}=1/2$.

\section{Superconformal field theories}
\label{Superconformal field theories}

The interest in superconformal field theories is due to the important restrictions imposed by supersymmetry. In the context of  supersymmetric conformal field theories not only there exist exactly solvable models but also there are quantities that can be exactly computed even without solving completely the theory. This includes dimensions of chiral operators, since they are connected with the $R$-charge, and central charges. In addition, as we will discuss in the next section, the analysis of conformal field theories with global symmetries becomes quickly numerically challenging; the presence of supersymmetry provides a relation between the coefficients of otherwise independent conformal blocks. This in turns makes the numerical procedure more powerful and precise, allowing to derive strong bounds for the case of $U(1)$ symmetry (corresponding in this case to the $R$-charge).\\
In this section we present the investigation of the four point functions of a complex scalar which is the lowest component field of a chiral superfied. In the analysis we will exploit the following properties of superconformal invariance:
\begin{itemize}
\item The operators of the theory are arranged in irreducible representations of the superconformal algebra. These, decomposed with respect to the usual conformal sub-algebra contain a finite number of conformal primary operators, but only one of them is a \emph{superconformal primary}. The others can be obtained acting with the supercharges, which play the role of raising operators in the superspace. Hence the contribution to the four point function of a superconformal representation will be the sum, with fixed coefficients, of a finite number of conformal blocks.
\item Unitarity bounds in presence of the superconformal algebra are more restrictive.
\item The OPE of two chiral (or a chiral and an anti-chiral) operators is constrained by superconformal invariance, thus the operators contributing to the four point function must obey to restrictions more stringent then those imposed only by unitarity. 
\end{itemize}

\subsection{Superconformal algebra}
\label{Superconformal algebra}

The superconformal algebra represents an extension to superspace of the ordinary conformal algebra. One of the possible way to define it is as the set of transformation acting on the superspace $(x^\m,\,\theta_\alpha,\,\bar\theta_{\dot\beta})$ that preserve the super-line element
\be
\	ds^2= (dx_\m + i\theta\sigma^\m \,d\bar\theta+i \bar\theta\bar\sigma^\m \,d\theta)^2
\ee
up to a conformal factor. Those consist in the ordinary conformal sub-algebra $P_\mu, \, M_{\mu\nu},\, K_\mu,\, D$, the supersymmetry generators $Q_\alpha,\, \bar Q_{\dot \alpha}$, the $R-$charge generator $A$ and two more fermionic generator $S_\alpha,\, \bar S_{\dot \alpha}$. Unless explicitly stated we follow the conventions of \cite{poland}.

\subsection{Representation of the Superconformal Algebra and unitarity bounds}
\label{Representation of the Superconformal Algebra and unitarity bounds}

Let us start describing qualitatively the structure of highest weight representations of the superconformal algebra and how they decompose with respect to the conformal sub-algebra. The lowest dimension state of the representation is called \emph{superconformal primary} and satisfies the condition
\be\label{superprimary}
\	[S_\alpha,\,O]= [\bar S_{\dot\alpha},\,O]=[K_\mu,\, O]=0\,,
\ee
The higher states of the representation can be obtained actioning with the raising operators $P_\m\,\,,Q_\alpha,\,\bar Q_{\dot\alpha}$. The effect of $P_\m$ has already been discussed and reproduces the ordinary descendants operators. The action of the supercharges instead can produces operators that are still conformal primaries but not superconformal primaries any more. For instance, given a field satisfying the condition (\ref{superprimary}) we have
\be
\	[K_\m,\, [Q_\alpha\,,O]]= i (\sigma_\mu)_{\alpha\dot\beta}\bar [S^{\dot \beta},O]=0 
\ee
implying that $ [Q_\alpha\,,O]$ is again a primary operator.  This means that a super field contains several primary operators, corresponding to different powers of $\theta,\,\bar\theta$. Notice that all the non-superconformal primaries appearing in a given representation have dimension higher than the dimension of the superconformal primary. This is by construction a consequence of the fact that the lowest state of the representation is defined as the state annihilated by all the operators that lower the dimension. On the other hand, the lowest state has not necessarily the minimal spin among the primaries. Indeed there can be primary operators with lower spin. Finally, the $R$-charge of the non-superconformal primary can be at maximum one unit larger or smaller than the one of the superconformal primary.\footnote{This because the expansion in Grassman variables ends at quadratic order.} 

An irreducible representation of the super conformal algebra with $\mathcal N=1$ is labelled by 4 numbers:
\be
\	\left(q\,,\bar q\,, j_1\,,j_2\right)
\ee
where $q\,,\bar q$ are related to the scaling dimension and the $R-$charge of the superconformal primary according to:
\be
\	\Delta=q+\bar q\,, \qquad \frac32 R=q-\bar q
\ee
The unitarity bounds read ( \cite{Dobrev:1985qv}):
\ba\label{susy unitarity bound}
\	 &&q\geq j_1+1\,, \qquad \bar q\geq j_2+1\,,\nn\\
	\Rightarrow && \Delta\geq \left|\frac32 R-j_1+j_2\right|+j_1+j_2+2\,.
\ea
The equality in the above relation is realized by the so called semi-conserved currents, for which it holds
\ba
\	\frac{3}{2}R=j_1-j_2\, \qquad \Delta=\frac{3}{2}R+2j_2+2=-\frac{3}{2}R+2j_1+2
\ea
If the superfield is also real its $R$-charge vanishes and $j_1=j_2$. These operators corresponds to traceless symmetric tensors of rank $l=2j_1=2j_2$. 
As in usual supersymmetry we can have multiplet shortening, with therefore a different structure of the unitarity bounds. Whenever one of the parameters $j$ is zero we have the condition
\be
\	\Delta\geq j+1
\ee
The equality this time identifies chiral (or antichiral) superfields:
\be
\	\bar q=j_2=0\,\qquad \Delta=\frac32R\geq j_1+1\,.
\ee
Finally the unit operator corresponds to $q=\bar q=j_1=j_2=0$.

\subsection{$\mathcal N=1$ OPE of Chiral Superfields }
\label{OPE of Chiral Superfields}

Before presenting the explicit form of the superconformal blocks let us review the OPE expansion for chiral fields. This will be crucial to understand what are the operators allowed to contribute to the four point function an will let us restrict the number of constraints to impose. Our analysis agrees with the latest version of \cite{poland}.\\
Consider now the OPE of a chiral superfields $\Phi(x,\theta,\bar\theta)$ with dimension $d=\frac{3}{2}R_\Phi$ with itself:
\be\label{eq:OPEphiphi}
\	\Phi\times \Phi 
\ee
According to the usual relation between the OPE coefficient and the three point function the above expression will contain the operator $O_I$ (here $I$ represents the space-time indices) if and only if the correlator
\be\label{susy3point}
\	\langle \Phi\Phi O_I^\dagger\rangle
\ee
is non vanishing. A complete characterization of the general form of three point function of generic superfields can be found in \cite{Osborn:1998qu}. For the case of two scalar superfields and a third spin-$l$ superfield, it can be written as
\ba
\	&&\langle \Phi(x_1,\theta_1,\bar\theta_1)\Phi(x_2,\theta_2,\bar\theta_2) O_I^\dagger(x_3,\theta_3,\bar\theta_3)\rangle=\frac{t_I(X_3,\Theta_3,\bar\Theta_3)}{x_{\bar31}^{2d}x_{\bar32}^{2d}}\,,\nn\\
\	&& x_{\bar i j}=x_{i-}+x_{j+}+2i\theta_j \sigma \bar\theta_i\,,\qquad x_{i\pm}=x\pm i\theta_i \sigma \bar\theta_i\,,\nn\\
\	&& X_3^\mu=-\frac12\frac{ x_{\bar 31}^\n x_{\bar 12}^\rho x_{\bar 23}^\gamma}{{x_{\bar31}^{2}x_{\bar32}^{2}}}\text{Tr}[\bar\sigma^\mu\sigma_\n\bar\sigma_\rho\sigma_\gamma]\,\qquad \theta_{ij}=\theta_i-\theta_j\nn\\
\	&& \Theta_3= i\frac{x_{\bar31}^\mu}{x_{\bar13}^2}\sigma_\mu\bar\theta_{31}-i\frac{x_{\bar32}^\mu}{x_{\bar23}^2}\sigma_\mu\bar\theta_{32}\,,\qquad  \bar\Theta_3= \Theta_3^\dagger\,.
\ea
where $t_I(X_3,\Theta_3,\bar\Theta_3)$ has to be determined. 
In addition, further restrictions must be imposed. First, the correct transformation properties under the superconformal group are realized if $t_I$ satisfies the homogeneity condition:
\ba\label{eq:hom cond}
\	t_I(\lambda \bar \lambda X_3,\lambda \Theta_3,\bar \lambda \bar\Theta_3)=\lambda^{2a} \bar \lambda^{2\bar a}t_I(X_3,\Theta_3,\bar\Theta_3)\,,\\
\	a=\frac{1}{3}\left(2q_O+\bar q_O-4d\right)\,,\qquad \bar a=\frac{1}{3}\left(2\bar q_O+ q_O-2d\right)\,.
\ea
Moreover, the chirality of $\Phi$ translates in the condition:
\ba
\	&&\bar D_{\dot\alpha}\frac{t_I(X_3,\Theta_3,\bar\Theta_3)}{x_{\bar31}^{2d}x_{\bar32}^{2d}}= \frac{\bar D_{\dot\alpha}t_I(X_3,\Theta_3,\bar\Theta_3)}{x_{\bar31}^{2d}x_{\bar32}^{2d}}\\
\	&&\phantom{\bar D_{\dot\alpha}\frac{t_I(X_3,\Theta_3,\bar\Theta_3)}{x_{\bar31}^{2d}x_{\bar32}^{2d}}}= -i \frac{1}{x_{\bar31}^{2d}x_{\bar32}^{2d}}\frac{(x_{\bar 13})^{\dot\alpha\alpha}}{x_{\bar31}}\left(\frac{\p}{\p \Theta_3^\alpha}-2i(\sigma^\mu\bar\Theta_3)_\alpha \frac{\p}{\p X_3^\m}\right)t_I(X_3,\Theta_3,\bar\Theta_3)\nn
\ea
which has the general solution:
\ba
\	&& t_I(X_3,\Theta_3,\bar\Theta_3)=t_I(\bar X_3,\bar\Theta_3)\,,\nn\\
\	&& \bar X_3 = X_3+2i \Theta \sigma \bar \Theta\,.
\ea
At this point we can look for a generic function of the above form that satisfies the homogeneity condition (\ref{eq:hom cond}):
\ba
\	t_I(\bar X_3,\bar\Theta_3)\sim \bar X_3^m \bar \Theta^n\,,\qquad n=0,1,2\,,\nn\\
\	2a=m\,,\qquad 2\bar a=n+m
\ea
A final constraint come from the invariance under exchange $1\leftrightarrow 2$, since the chiral operators are the same. This requirement translate in the invariance under $\bar X_3\rightarrow -X_3\,, \bar \Theta_3\rightarrow -\Theta_3\,, $.
We found three possible solutions consistent with all the constraints:
\ba\label{solutsusyOPE}
\	&&\text{constant} ,\qquad  \phantom{--------}R_O=\frac{4}{3}d\,,\quad \D_O=2d \,,\quad l=0\, \nn\\ 
	&& \bar\Theta_3  \bar X_3^{\m_1}...\bar X_3^{\m_l}\,,\qquad \phantom{------}R_O=\frac{4}{3}d-1\,,\quad \D_O=2d+l+\frac12\,,\quad \text{l odd}\nn\\
	&&  \bar\Theta_3^2  \bar X_3^{\Delta_O-2d-1-l} \bar X_3^{\m_1}...\bar X_3^{\m_l}\,,\qquad R_O=\frac{4}{3}d-2\,,\quad \Delta_O\geq |2d-3|+l+2\, \quad \text{l even}.
\ea
 We finally conclude that only the following superconformal primary can be present in the OPE (\ref{eq:OPEphiphi}):
\begin{itemize}
\item Chiral operator with $\Delta=\frac{3}{2}R=2d$ and $l=0$. We will denote it $\Phi^2$.
\item Non-Chiral operators with $j_1-j_2=1/2$, $l=2j_2$ odd, $R=\frac{4}{3}d-1$ and dimension\footnote{Eq. (\ref{comparisonwithpoland}) agrees with the results of \cite{poland} once one redefines $l\rightarrow l-1$.}
\be\label{comparisonwithpoland}
\	\Delta_O= 2d+l+\frac12\,.
\ee
\item Non-Chiral operators with $l=2j_1=2j_2$ even, $R=\frac{4}{3}d-2$ and dimension
\be
\	\Delta_O\geq | 2d-3|+l+2\,.
\ee
\end{itemize}
We should also stress that none of the expressions in (\ref{solutsusyOPE}) can be further expanded in $\Theta_3\sigma\bar\Theta_3$ (which is hidden inside $\bar X_3$)\footnote{While in the third line of (\ref{solutsusyOPE})  the expansion is trivial since $ \bar\Theta_3^2$ already saturates the antisymmetric properties of Grassman variables, the second line of requires a more delicate analysis.}.
This fact has a very crucial consequence: if we take $\theta_1=\theta_2=0$ in (\ref{susy3point}) and we expand in $\theta_3,\bar \theta_3$ we obtain a series in the Grassman variables whose coefficients are three point function of two scalars with the lowest component of $O_I$ or one of its super-descendants. On the other hand, the expressions in (\ref{solutsusyOPE}) contain only one term each. This means that there is only one operator in the superconformal representation, with the correct $R$-charge, that have a non vanishing three point function with $\phi\phi$. \\
From the above results we can infer what are the operators appearing in the OPE of the lowest component of the chiral superfield $\Phi$ with itself. Call $\phi$ this scalar field with dimension $d$ than $\phi\times\phi$ receive contributions from the lowest component of
\begin{itemize}
\item Chiral superfield $\Phi^2$, call it $\phi^2$.
\item super-descendant $(\sigma^\mu)^{\beta\dot\alpha} \bar Q_{\dot\alpha} O_{\beta}^{\m_1...\m_l}$. In this case the operator contributing to the OPE  $\phi\times\phi$ is a $(l+1)$-rank tensor, $(l+1)$ even and:
\be\label{constraint1}
\	R_{\bar QO}=\frac{4}{3}d\,,\qquad \Delta_{\bar QO} = 2d+(l+1)\,.
\ee
Notice that in the free case, $d=1$, the above operators are precisely twist-2 operators with even spin, as expected from the expansion of the four point function $\langle\phi\phi\phi^\dagger\phi^\dagger\rangle$ in the s-channel in the non supersymmetric case.

\item super-descendant $ \bar Q^2 O$. In this case the operator contributing to the OPE  $\phi\times\phi$ is a $l$-rank tensor, $l$ even and:
\be\label{constraint2}
\	R_{\bar Q^2O}=\frac{4}{3}d\,,\qquad \Delta_{\bar Q^2O} \geq |2d-3|+l+3\,.
\ee
\end{itemize}
We will also need the general structure of the OPE 
\be\label{eq:OPEphiphidagger}
\	\Phi\times \Phi^\dagger
\ee
As before we need to study the form of the correlator
\be
\	\langle \Phi\Phi^\dagger O_I^\dagger\rangle
\ee
allowed by superconformal symmetry. Starting from the general form (\cite{Osborn:1998qu})
 
\ba
\	&&\langle \Phi(x_1,\theta_1,\bar\theta_1)\Phi^\dagger(x_2,\theta_2,\bar\theta_2) O_I^\dagger(x_3,\theta_3,\bar\theta_3)\rangle=\frac{t_I(X_3,\Theta_3,\bar\Theta_3)}{x_{\bar31}^{2d}x_{\bar32}^{2d}}\,,\nn\\
\	&& t_I(\lambda \bar \lambda X_3,\lambda \Theta_3,\bar \lambda \bar\Theta_3)=\lambda^{2a} \bar \lambda^{2\bar a}t_I(X_3,\Theta_3,\bar\Theta_3)\,,\\
\	&& a=\frac{1}{3}\left(2q_O+\bar q_O-3d\right)\,,\qquad \bar a=\frac{1}{3}\left(2\bar q_O+ q_O-3d\right)\,,
\ea
we can supplement the chirality condition of the $\Phi$, imposing $ t_I(X_3,\Theta_3,\bar\Theta_3)= t_I(\bar X_3,\bar\Theta_3)$, with the anti-chirality condition of $\Phi^\dagger$:
\ba
\	0= D_{\alpha}\frac{t_I(\bar X_3,\bar\Theta_3)}{x_{\bar31}^{2d}x_{\bar32}^{2d}}=-i \frac{1}{x_{\bar31}^{2d}x_{\bar32}^{2d}}\frac{(x_{ 2\bar3})^{\dot\alpha\alpha}}{x_{\bar2 3}^2}\frac{\p}{\p\bar \Theta_3^{\dot\alpha}} t_I(\bar X_3,\bar\Theta_3)
\ea
The only possibility is to have $t^I$ depending only on $\bar X^\mu_3$. In the end the only structure admitted for the three point function is
\ba\label{PhiPhiO}
\	&&  \langle \Phi(x_1,\theta_1,\bar\theta_1)\Phi^\dagger(x_2,\theta_2,\bar\theta_2) O_I^\dagger x_3,\theta_3,\bar\theta_3\rangle=\frac{C_{\Phi\Phi^\dagger O}}{x_{\bar31}^{2d}x_{\bar32}^{2d}} \left( \bar X_3^{\Delta_O-2d-l} \bar X_3^{\m_1}...\bar X_3^{\m_l}-\text{traces}\right)\,,\nn\\
\	 && R_O=0\,, \quad \text{l integer}\,,\quad \Delta_O\geq l+2.
\ea
where we used the unitarity bound (\ref{susy unitarity bound}) for traceless rank$-l$ operator with vanishing $R$-charge.\\
Setting $\theta_{1,2}=\bar\theta_{1,2}=0$ in the above expression one can therefore reduce to the three point function of the lowest component $\phi\,\phi^\dagger$, with the a third superfield $O$. Expanding in $\theta_3,\bar\theta_3$ and matching the corresponding terms one can finally extract the contribution of the superconformal primary (the zeroth order term) and those of the other conformal primaries contained in the superfield $O$. Clearly only operators with vanishing $R$-charge will contribute to the three point function, since only those operators can appear in the $\phi\times\phi^\dagger$ OPE. This operators corresponds to the term linear and quadratic in the combination $\theta \sigma^\mu \bar\theta$ and therefore corresponds to spin and dimension $(\Delta+1,l+1),\,(\Delta+1,l-1)\,,$ and $(\Delta+2,l)$. Moreover the relative coefficients are totally fixed by supersymmetry and the only unknown quantity is the overall constant $C_{\Phi\Phi^\dagger O}$ appearing in the three point function (\ref{PhiPhiO}). Those coefficients have been computed explicitly  in \cite{poland}.\\

\subsection{$\mathcal N=1$ Superconformal blocks}
\label{Superconformal blocks}

According to the discussion of the previous section we can now infer the most general  form of the four point function
\be
\	\langle \phi(x_1)\phi^\dagger(x_2)\phi(x_3)\phi^\dagger(x_4) \rangle\,,
\ee
allowed by superconformal invariance. Here $\phi$ is the lowest component of a chiral superfield with dimension $d$. Generically the above correlator can be expressed in a sum of conformal blocks. 
We notice that there are two alternative and equivalent ways to express the above four point function. The s-channel corresponds to take the OPE $\phi(x_1)\times\phi^\dagger(x_2)$ and $\phi(x_3)\times\phi^\dagger(x_4)$. These OPE's contain operators with vanishing $R$-charge\footnote{In supersymmetry the role of the $U(1)$ symmetry is played by the $R$-charge. Although this symmetry is not properly a global symmetry in superspace, it commutes with the conformal sub-algebra.} and integer spin, even and odd. Moreover the contribution of conformal primaries belonging to the same superconformal representation are related by known coefficients and can be grouped in \emph{superconformal blocks}:
\be\label{susypartialwave}
\	\langle
	 \bcontraction[1ex]{}{\phi(x_1)}{}{\phi^\dagger(x_2)}
	 \bcontraction[1ex]{\phi(x_1)\phi^\dagger(x_2)}{\phi(x_3)}{}{\phi^\dagger(x_4)}  
	 \phi(x_1)\phi^\dagger(x_2)\phi(x_3)\phi^\dagger(x_4) 
	 \rangle = \frac{1}{x_{12}^{2d}x_{34}^{2d}}\left(1+ \sum_{\Delta\geq l+2}  p_{\Delta,l} \mathcal G_{\Delta,l}(u,v) \right)\,,
\ee
where we defined 
\ba\label{susyCB}
\	&&\mathcal G_{\Delta,l}(u,v)= g_{\Delta,l}(u,v)+\frac{(\Delta+l)}{4(\Delta+l+1)}g_{\Delta+1,l+1}(u,v)+\frac{(\Delta-l-2)}{4(\Delta-l-1)}g_{\Delta+1,l-1}(u,v)\nn\\
\	&&\phantom{-------}+\frac{(\Delta+l)(\Delta-l-2)}{16(\Delta+l+1)(\Delta-l-1)}g_{\Delta+2,l}(u,v)\,.
\ea
As discussed at the end of the previous section the superconformal blocks encodes the contribution of four operators with dimension and spin $(\Delta+1,l+1),\,(\Delta+1,l-1)\,,$ and $(\Delta+2,l)$. The difference from \cite{poland} is only due to a different normalization of the ordinary conformal block: we have removed the pre-factor $(-1/2)^{l}$. \\
Performing the OPE expansion in the $t$-channel produces again an expansion in terms of superconformal blocks, with a different dependence on coordinates:
\ba
\	&&\langle
	 \bcontraction[1ex]{}{\phi(x_1)}{\phi^\dagger(x_2)\phi(x_3)}{\phi^\dagger(x_4)}
	 \bcontraction[2ex]{\phi(x_1)}{\phi^\dagger(x_2)}{}{\phi(x_3)}  
	 \phi(x_1)\phi^\dagger(x_2)\phi(x_3)\phi^\dagger(x_4) 
	 \rangle = \frac{1}{x_{14}^{2d}x_{23}^{2d}}\left(1+ \sum_{\Delta\geq l+2} p_{\Delta,l} \mathcal G_{\Delta,l}\left(v,u\right)\right)
\ea
Equating the two expansions we get one of the crossing symmetry constraint for $U(1)$-invariant theories, except that now only superconformal blocks enter in the sum rule:
\ba\label{first sum rule}
\	&&\sum_{\Delta\geq l+2} p_{\Delta,l} \mathcal{F}_{\Delta,l}=1\,,\\
\	&&  \mathcal{F}_{\Delta,l}=\frac{u^{-d}\mathcal G_{\Delta,l}\left(u,v \right)-v^{-d}\mathcal G_{\Delta,l}\left(v,u\right)}{v^{-d}-u^{-d}}\nn
\ea
Additional constraints can be derived as usual considering the $s-$ and $t-$ channel expansions of a different four point function. Let us start from the following decomposition
\ba
\	&&\langle
	 \bcontraction[1ex]{}{\phi(x_1)}{}{\phi^\dagger(x_2)}
	 \bcontraction[1ex]{\phi(x_1)\phi^\dagger(x_2)}{\phi^\dagger(x_3)}{}{\phi(x_4)}  
	 \phi(x_1)\phi^\dagger(x_2)\phi^\dagger(x_3)\phi(x_4) 
	 \rangle = \frac{1}{x_{12}^{2d}x_{34}^{2d}}\left(1+ \sum_{\Delta\geq l+2}  p_{\Delta,l} \mathcal G_{\Delta,l}\left(\frac{u}{v},\frac{1}{v}\right) \right)\nn\\
	 &&\phantom{\langle
	 \bcontraction[1ex]{}{\phi(x_1)}{\phi^\dagger(x_2)\phi(x_3)}{\phi^\dagger(x_4)}
	 \bcontraction[2ex]{\phi(x_1)}{\phi^\dagger(x_2)}{}{\phi(x_3)}  
	 \phi(x_1)\phi^\dagger(x_2)\phi(x_3)\phi^\dagger(x_4) 
	 \rangle }= \frac{1}{x_{12}^{2d}x_{34}^{2d}}  \left(1+ \sum_{\Delta\geq l+2}(-1)^l p_{\Delta,l} \widetilde{\mathcal G}_{\Delta,l}\left(u,v \right)\right)\,,
\ea
where we have used the known property of the conformal blocks 
\be\label{CBparity}
g_{\Delta,l}(u,v)=(-)^{l}g_{\Delta,l}(u/v,1/v).
\ee
to flip the signs of odd spins and we have defined
\ba
\	&&\widetilde{\mathcal G}_{\Delta,l}(u,v)= g_{\Delta,l}(u,v)-\frac{(\Delta+l)}{4(\Delta+l+1)}g_{\Delta+1,l+1}(u,v)-\frac{(\Delta-l-2)}{4(\Delta-l-1)}g_{\Delta+1,l-1}(u,v)\nn\\
\	&&\phantom{-------}+\frac{(\Delta+l)(\Delta-l-2)}{16(\Delta+l+1)(\Delta-l-1)}g_{\Delta+2,l}(u,v)\,.
\ea
The $t-$channel decomposition corresponds to take OPE's $\phi\times\phi$ and its conjugate. These OPE's contains operators with $R-$charge twice the $R$-charge of $\phi$. As discussed in the previous section only one primary operator per representation contributes, that is to say superconformal blocks in this channel reduce to ordinary conformal blocks. Moreover the dimensions of the primary operators are subject to the constraint (\ref{constraint1}) or the bound (\ref{constraint2}). Hence: 
\ba
	&&\langle
	 \bcontraction[1ex]{}{\phi(x_1)}{\phi^\dagger(x_2)\phi^\dagger(x_3)}{\phi(x_4)}
	 \bcontraction[2ex]{\phi(x_1)}{\phi^\dagger(x_2)}{}{\phi^\dagger(x_3)}  
	 \phi(x_1)\phi^\dagger(x_2)\phi^\dagger(x_3)\phi(x_4) 
	 \rangle = \frac{1}{x_{14}^{2d}x_{23}^{2d}}\left( \sum_{l \text{ even}} p^{R=2}_{2d+l,l} \,g_{2d+l,l}\left(v,u\right)+ \sum_{\substack{\Delta\geq |2d-3|+l+3\\ l\text{even} } }p^{R=2}_{2d+l,l} \, g_{\Delta,l}\left(v,u\right)\right)\nn\\
\ea
Notice that the first term in parenthesis, when $l=0$, contains the contribution of the chiral operator $\Phi^2$. Equating the two expansion we finally get two more sum rules:
\ba\label{other sum rule}
\	&&\sum_{\Delta\geq l+2} p_{\Delta,l} (-1)^l \widetilde{\mathcal{F}}_{\Delta,l}\,\,+\sum_{\substack{\Delta=2d+l\\ \Delta\geq |2d-3|+l+3\\ l \text{ even}}} p_{\Delta,l}^{R=2} F_{\Delta,l}=1\,,\\
\	&&\sum_{\Delta\geq l+2} p_{\Delta,l} (-1)^l \widetilde{\mathcal{H}}_{\Delta,l}\,\,-\sum_{\substack{\Delta=2d+l\\ \Delta\geq |2d-3|+l+3\\ l \text{ even}}} p_{\Delta,l}^{R=2} H_{\Delta,l}=-1\,,\nn
\ea
where $F,H$ are defined in (\ref{eq:F}, \ref{eq:H}) and
\ba
\	&&  \widetilde{\mathcal{F}}_{\Delta,l}=\frac{u^{-d} \widetilde{\mathcal G}_{\Delta,l}\left(u,v \right)-v^{-d} \widetilde{\mathcal G}_{\Delta,l}\left(v,u\right)}{v^{-d}-u^{-d}}\,\nn\\
\	&&  \widetilde{\mathcal{H}}_{\Delta,l}=\frac{u^{-d} \widetilde{\mathcal G}_{\Delta,l}\left(u,v \right)+v^{-d} \widetilde{\mathcal G}_{\Delta,l}\left(v,u\right)}{v^{-d}+u^{-d}}
\ea
The sum rule (\ref{first sum rule}) has been used in \cite{poland} to derive bounds on the dimension of the lowest dimension scalar operator entering in in the OPE $\phi\times\phi^\dagger$. In the next chapter we will reproduce their results and show that the additional use of (\ref{other sum rule}) allows the extraction of stronger results with less numerical effort.

\subsubsection{Free Theory}

In order to check that we didn't miss any sign we can compute the expansion in superconformal blocks and verity that the obtained spectrum and coefficient solve the vectorial sum rule. A theory of a complex free scalar can be trivially made supersymmetric adding a free Weyl fermion. Since both the fields are free there is no modification of the OPE's nor of the correlation functions. Hence we know that the OPE of a complex field contains only twist-2 operators (a $U(1)$ theory is the same as a $SO(2)$ theory). From the analysis of $U(1)$ theories we also know the decomposition of the four point function in terms of conformal blocks. On the other hand here we are interested in the superconformal block decomposition:
\[
\left\langle \phi(x_{1})\phi^{\dagger}(x_{2})\phi(x_{3})\phi^{\dagger}%
(x_{4})\right\rangle =\frac{1}{x_{12}^{2}x_{34}^{2}}  \left(1+\sum_{l}p_{l+2,l} \left( g_{l+2,l}(u,v)+\frac{2l+2}{2l+3} g_{l+3,l+1}(u,v)\right)\right)
\]
In the above expansion we have collected together the contribution to the four point function of an entire supermultiplet. It easy to see that the coefficients satisfy
\begin{equation}\label{psusy}
p_{l+2,l}=\frac{(l!)^{2}}{(2l)!}\frac{l+1}{2l+1}%
\end{equation}
On the other hand, the coefficient $p_{\Delta,l}^{R=2}$ are exactly the same as in the free complex scalar:
\be
\	p_{l+2,l}^{R=2}=(1+(-1)^l)\frac{(l!)^2}{(2l)!}
\ee
One can numerically verify the fast convergence of the three sum rules (\ref{first sum rule}), (\ref{other sum rule}) in the interval $b=0\, ,a\in [-0.5,\, 0.5]$.

\section{Bounds and numerical results}
\label{Bounds and numerical results}

As extensively discussed in \cite{r1,r2,r3,poland} crossing symmetry constraint are not compatible with arbitrary spectra of operators nor with arbitrary large OPE coefficients. In particular, bounds on the dimension $\Delta_{\text{min}}$ of the first scalar operator entering the OPE of two scalar fields can be extracted. \\
The method introduced in \cite{r1} exploits the language of linear functionals. Let us first review the logic for the case of a single sum rule.
A linear functional $\Lambda$ on space of functions of two variables $\{F(a,b) \}$ is given by
\begin{equation}
\Lambda=	\sum_{m,n}\lambda_{2m,2n}\partial^{2m}_a\partial^{2m}_b\,,\qquad 
\Lambda(F)=\sum_{\mathcal{B}}\lambda_{2m,2n}F^{(2m,2n)}, \label{fun0}%
\end{equation}
where $\lambda_{2m,2n}$ are some fixed numbers characterizing the functional and all derivatives are evaluated at $a=b=0$. The variables $a,b$ are related to $z,\bar z$ via:
\be
\	z=\frac12+a+b\,,\qquad \bar z=\frac12+a-b\,.
\ee
An upper bound on $\Delta_{\text{min}}$ can be extracted according to the following procedure. Assume thus that for certain fixed $d$ and $\Delta_\text{min}%
,$ we manage to find a linear functional of this form such that
(\emph{\textquotedblleft positivity property\textquotedblright})%
\ba\label{posprop}
\Lambda\lbrack F_{d,\Delta,l}]\geq0\text{ }  &&  \text{for all }\Delta
\geq\Delta_{\min}\,(l=0)\label{eq-ineq}\\
\text{and }  &&  \text{for all }\Delta\geq l+2\,(l=2,4,6\ldots)\,,\\
\Lambda[1] \leq 0 && \,\nonumber
\ea
Moreover, assume that all but a finite number of these inequalities
are actually strict: $\Lambda\lbrack F]>0.$ Then the sum rule cannot
be satisfied, and such a spectrum, corresponding to a putative OPE
$\phi\times\phi$, is ruled out.\\
The proof uses the above \textquotedblleft positivity
argument\textquotedblright. Since $\Lambda\lbrack1]\leq0,$ the positivity
property implies that only those primaries for which $\Lambda\lbrack F]=0$
would be allowed to appear in the RHS of the sum rule with nonzero
coefficients. By assumption, there are at most a finite number of such
primaries. However, as noted in \cite{r1}
 finitely many terms
can never satisfy the sum rule globally, because of the behavior near $z=0,1.$\\

\subsection{Results for CFT's with global symmetries}

\label{Results for CFT's with global symmetries:dim}

As discussed in Sec.\ref{CFT's with Global symmetries}, the presence of global symmetries in the CFT implies that the operators appearing in the OPE of two scalar field can be classified according to their representation. Moreover, as shown in \cite{r4}, the number of sum rules arising from crossing symmetry constraints is equal in number to the number of structures. We recall that by \emph{structure} we denote the contribution to the four point function of all the operators with the same spin parity belonging to the same representation of the global symmetry. In the $SO(N)$ case there are three structures: singlet and symmetric traceless with even spin, antisymmetric with odd spin. Correspondingly we have three sum rules.\\
In order to extract information from the system of sum rules we adopt again the method of linear functionals. Here we denote $\Lambda$ a functional defined on the space of vector functions. Let us review the procedure for the simple case of $SO(N)$. Recalling eq. \ref{eq:son-vect} :
\be\label{vecsumrule}
\sum p_{\Delta,l}^{S}
\underbrace{
\left(
\begin{array}
[c]{c}%
0\\
F_{d,\Delta,l}\\
H_{d,\Delta,l}%
\end{array}
\right)}_{\vec V^S_{\Delta,l}}
 +\sum p_{\Delta,l}^{T}
 \underbrace{
 \left(
\begin{array}
[c]{c}%
F_{d,\Delta,l}\\
\left(  1-\frac{2}{N}\right)  F_{d,\Delta,l}\\
-\left(  1+\frac{2}{N}\right)  H_{d,\Delta,l}%
\end{array}
\right)}_{\vec V^T_{\Delta,l}}
  +\sum p_{\Delta,l}^{A}
 \underbrace{ 
  \left(
\begin{array}
[c]{c}%
-F_{d,\Delta,l}\\
F_{d,\Delta,l}\\
-H_{d,\Delta,l}%
\end{array}
\right) }_{\vec V^A_{\Delta,l}}
 =
\underbrace{
 \left(
\begin{array}
[c]{c}%
0\\
1\\
-1
\end{array}
\right)}_{\vec V^{RHS}}
\ee %
we need a functional defined on the vectors function $\vec V_i$. Let us define $N_{der}$ the highest order of derivatives included in the functional. Such a  linear functional can be parametrized as follows. Given a vector of functions of two variables $\vec v=(v_1(a,b),\,v_2(a,b)\,v_3(a,b))$ we define
\be\label{vec funct}
\	\Lambda[\vec v] = \sum_{j=1}^3 \sum_{n,m=0}^{N_{der}} c_j^{2n,2m} v_j^{(2m,2n)}(0,0)\,.
\ee
We immediately notice that, at the same order of truncation, the number of coefficients needed to parametrize a linear functional in the $SO(N)$ case are the triple of the non-symmetric case. More in general, calling $N_{structures}$ the number of structure and sum rules, the number of coefficients will be
\be
\	\frac{(N_{der}+2)(N_{der}+4)}{8} \times N_{structures}
\ee
The degree of numerical complexity is therefore enhanced. However this not the only complication: even the number of constraints that must be satisfied increases proportionally to  $N_{structures}$. \\
The functional $\Lambda$  must indeed satisfy the suitable generalized positivity property. Suppose for instance that we are interested in extracting a bound on the smallest dimension scalar singlet operator contributing to the sum rule; then we have to look for a functional subjected to
\ba\label{posSON}
\Lambda\lbrack \vec V_{d,\Delta,l}^S]\geq0\,,\phantom{----. }\,\,  &&  \text{for all }\Delta
\geq\Delta_{\min}\,(l=0)\\
\Lambda\lbrack \vec V_{d,\Delta,l}^T]\geq0\,,\phantom{----. }\,\,  &&  \text{for all }\Delta
\geq 1\,(l=0)\nn\\
\	\Lambda\lbrack \vec V_{d,\Delta,l}^i]\geq0\text{ },\,\, i=S,T,  && \text{for all }\Delta\geq l+2\,(l=2,4,6\ldots)\,.\nn\\
\	\Lambda\lbrack \vec V_{d,\Delta,l}^A]\geq0\,,\phantom{----. }\,\,   && \text{for all }\Delta\geq l+2\,(l=1,3,5\ldots)\,.\nn\\
\	\Lambda\lbrack \vec V^{RHS}]\leq 0\,.\phantom{----. } &&  
\ea
As mentioned, generically the number of constraints is multiplied by $N_{structures}$ with respect to the non-symmetric case. Notice that in the scalar symmetric traceless structure ($\vec V^T,\, l=0 $) we impose the positivity on all the $\Delta\geq 1$ since this is the constraint imposed by unitarity \cite{mack}. \\
\begin{figure}[h]
\begin{center}
\includegraphics[scale=0.4]{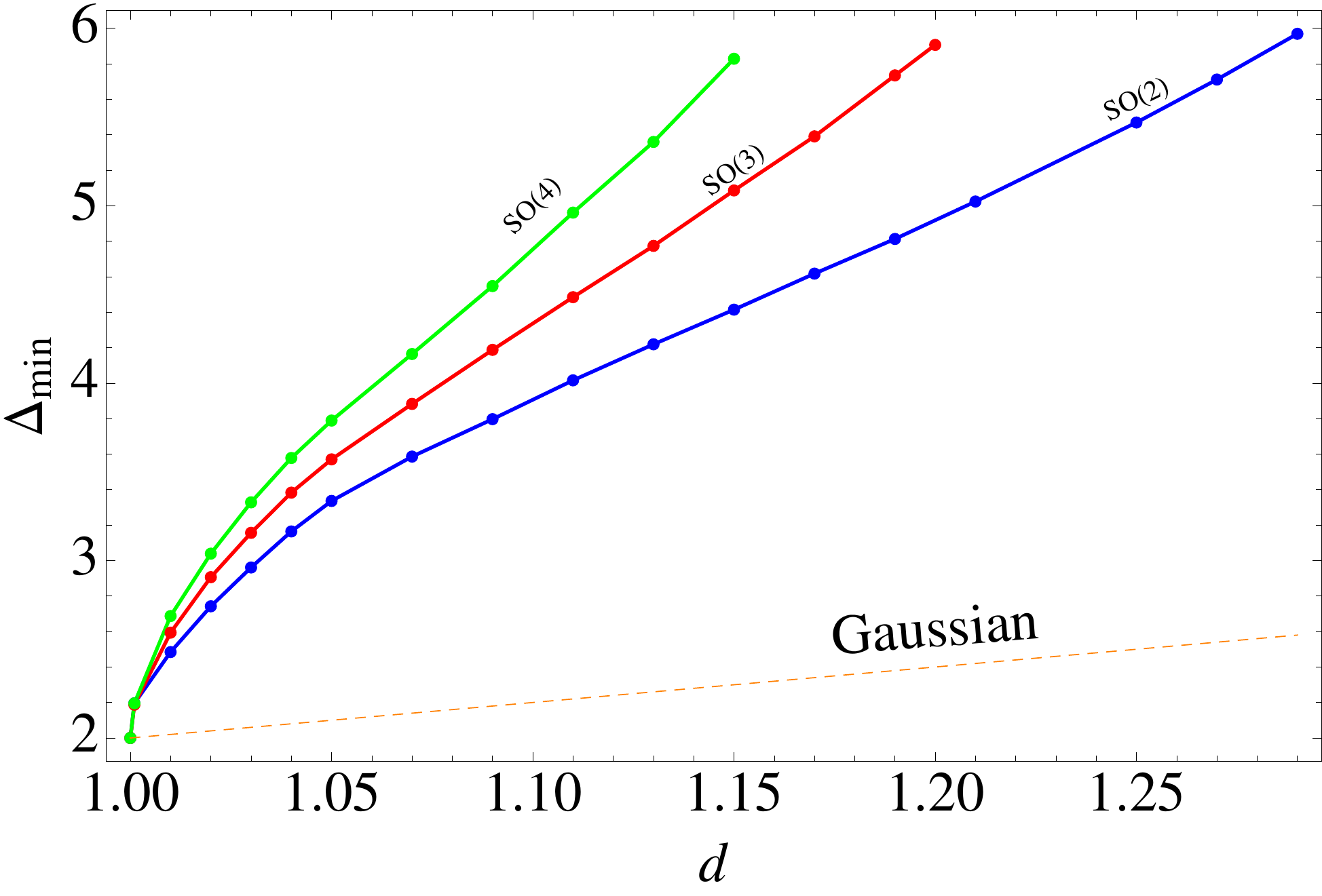}%
\caption[Bounds on dimension of scalar singlets in $SO(N)$ for $N=2,3,4$, ]{\footnotesize\textit{Bound for the smallest dimension of a scalar operators singlet under a global $SO(N)$ symmetries. The bounds corresponds, from the strongest to the weaker, to $SO(N),\,N=2,3,4$ and have been computed with 4 derivatives.
The line is an interpolation between the points where the bound has been computed exactly. We assume as usual a smooth interpolation.}}%
\label{figN}
\end{center}
\end{figure}
In \cite{r4} the existence of a bound was proved and few numerical results are provided. Here we push further the numerical calculations. The results for $N_{der}=4$ are shown in Fig. \ref{figN} for several theories. We observe that the bound gets weaker as $N$ increases. 

\subsubsection{$SO(4)$ and Conformal Technicolor}
\label{$SO(4)$ and Conformal Technicolor}

\begin{figure}[h]
\begin{center}
\includegraphics[scale=0.4]{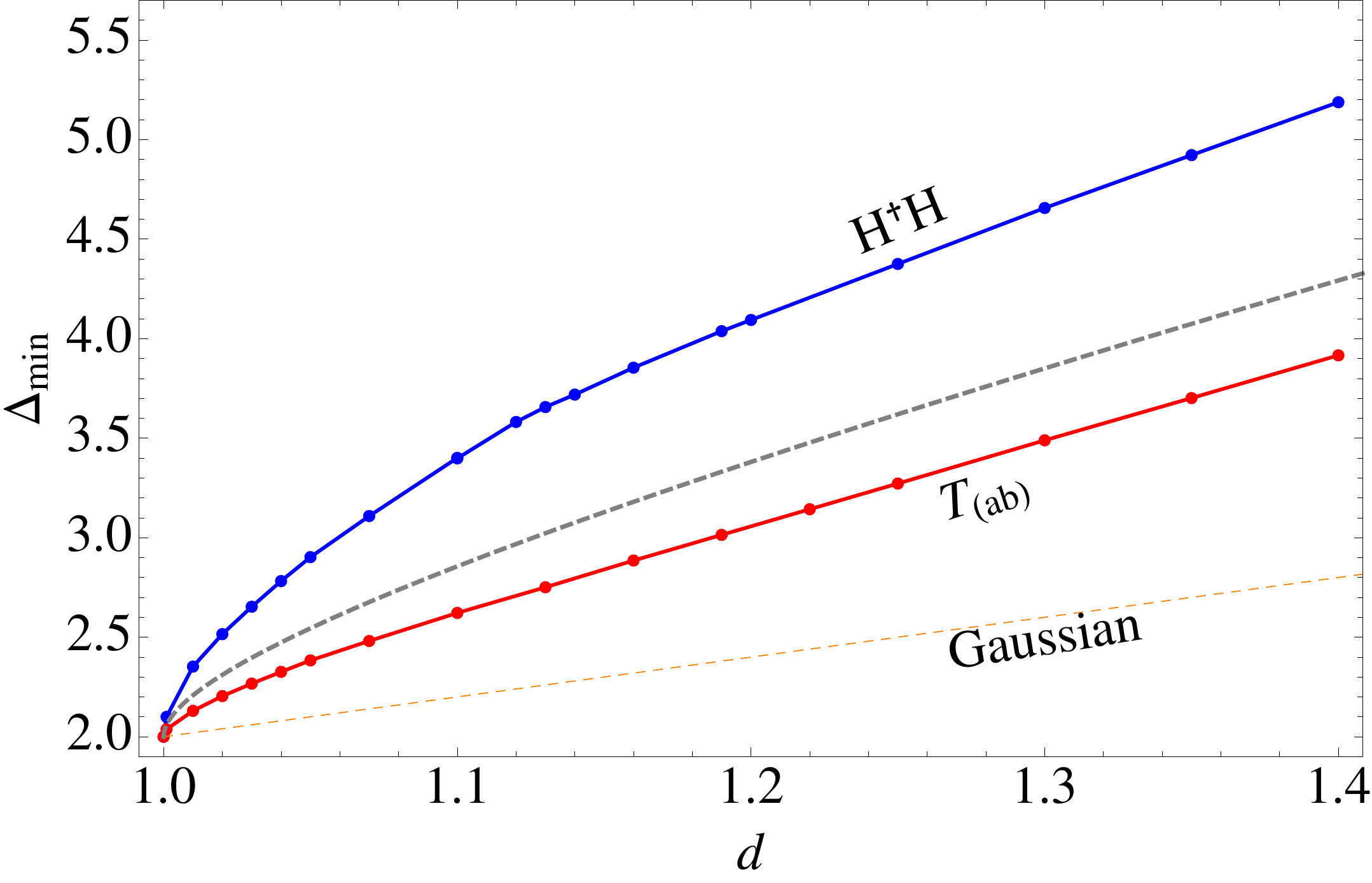}%
\caption[Bounds on the dimension of a scalar singlet and triplet in $SO(4)$.]{\footnotesize\textit{Bounds for the smallest dimension operators appearing in the OPE of two scalar fields transforming under the fundamental representation of a global $SO(4)$. The weaker bound (blue line) corresponds to scalar operators neutral under $SO(4)$. The strongest bound (red) refers to scalar operators transforming as a symmetric traceless tensor. Again we assumed a smooth interpolation between the points where the bound has been computed exactly.}}%
\label{SONsinglettriplet}
\end{center}
\end{figure}

The phenomenological motivation that motivated the original work \cite{r1} concerned what is the allowed separation between the dimension of a scalar field, identified with the Higgs $H$, and the dimension of the first scalar operator  appearing in its OPE, identified with the Higgs mass term $H^\dagger H$. A realistic scenario of BSM, where the Higgs is an operator of a strongly interacting CFT , must be able to accommodate a custodial $SO(4)$ symmetry in the strong sector, under which the Higgs field transforms as a fundamental representation while its mass term is a singlet\footnote{Notice that part of the custodial symmetry is gauged by the SM gauge group, hence if we want to include a CFT operator in the SM Lagrangian this has to be a singlet under custodial symmetry.}. As consequence we cannot directly apply stringent results of \cite{r2} to Conformal Technicolor, since the bounds produced in \cite{r1,r2} refer to the smallest operator appearing in the OPE without informations concerning its transformation properties under the global symmetry. \\
Using the formalism developed in this section we are now able to distinguish between different structures. \\
Recalling the OPE of a field transforming in the fundamental of $SO(4)$, call it $h_a$\footnote{In standard notation $h_a$ is related to H by $H=\frac1{\sqrt2}{h_1+ih_2\choose h_3+ih_4}$},
\begin{equation}
h_{a}(x)h_{b}(0)\sim\frac{1}{|x|^{2d}}\left(  \delta_{ab}\mathds 1+C_{S}%
|x|^{\Delta_{S}}\delta_{ab}\,(H^{\dagger}H)(0)+C_{T}|x|^{\Delta_{T}%
}\mathcal{T}_{(ab)}(0)+C_{J}|x|^{2}x^{\mu}J_{\mu}^{[ab]}(0)+\ldots\right)  \,.
\label{eq:OPEglobal}%
\end{equation}
we can extract a bound for the dimension $\Delta_S$ of the singlet operator and an independent bound on the dimension $\Delta_T$ on the symmetric traceless operator $T_{(ab)}$. The former makes use of the of the positivity properties as in (\ref{posSON}), while the latter uses similar constraints, reversing the role of $S$ and $T$, namely:
 \ba\label{posSONT}
\Lambda\lbrack \vec V_{d,\Delta,l}^T]\geq0\,,\phantom{----. }\,\,  &&  \text{for all }\Delta
\geq\Delta_{\min}\,(l=0)\\
\Lambda\lbrack \vec V_{d,\Delta,l}^S]\geq0\,,\phantom{----. }\,\,  &&  \text{for all }\Delta
\geq 1\,(l=0)\nn\\
\	\Lambda\lbrack \vec V_{d,\Delta,l}^i]\geq0\text{ },\,\, i=S,T,  && \text{for all }\Delta\geq l+2\,(l=2,4,6\ldots)\,.\nn\\
\	\Lambda\lbrack \vec V_{d,\Delta,l}^A]\geq0\,,\phantom{----. }\,\,   && \text{for all }\Delta\geq l+2\,(l=1,3,5\ldots)\,.\nn\\
\	\Lambda\lbrack \vec V^{RHS}]\leq 0\,.\phantom{----. } &&  
\ea
The extracted bounds are shown in Fig. \ref{SONsinglettriplet}, for $N_{der}=6$, and compared with the correspondent bound for non-symmetric theories computed with the same number of derivatives. The bound on $\Delta_T$ is the strongest one. This results points towards the generic expectation that the singlet operators should have higher dimension, confirming the explicit calculation for $O(N)$ models in $4-\epsilon$ dimensions (\cite{r1}). However we can't make a general statement about whether $\Delta_S-\Delta_T \gtrless 0$.\\
A detailed discussion of  the requirements of Conformal Technicolor (\cite{luty}) goes beyond the purposes of this work. \\ 
Indeed, to make a quantitative comparison between the needed pattern of field dimensions and our bound, assumptions on the physics of flavor must be made. \\
At the present stage theories containing clever assumptions on the flavor structure are allowed to live in sizable region of the parameter space $(d,\Delta_S)$, while more conservative ones are restricted to a small corner close to $d\sim1.2$. We should stress that increasing the numerical power in the non-symmetric case produced an improvement of 30-50\% in passing form $N_{der}=6$ to $N_{der}=18$. If a similar amelioration is repeated only theories exhibiting a clever flavor structure would remain unruled out, although severely constrained, while the others could not not be realized in a unitary CFT.

\subsection{Results for supersymmetric theories}
\label{Results for supersymmetric theories:dim}

Superconformal field theories theories represent another interesting class of theories where we can apply our formalism. Suppose we are given a superconformal field theory containing a chiral scalar field $\Phi$, the lowest component of which is a scalar complex field of dimension $d$. The crossing symmetry constraints arising from the four point function of $\langle\phi\phi^\dagger\phi\phi^\dagger\rangle$ have been revised in Sec. \ref{Superconformal blocks} and have the a similar structure of the vectorial sum rule (\ref{vecsumrule}). 
\begin{figure}[h]
\begin{center}
\includegraphics[scale=0.395]{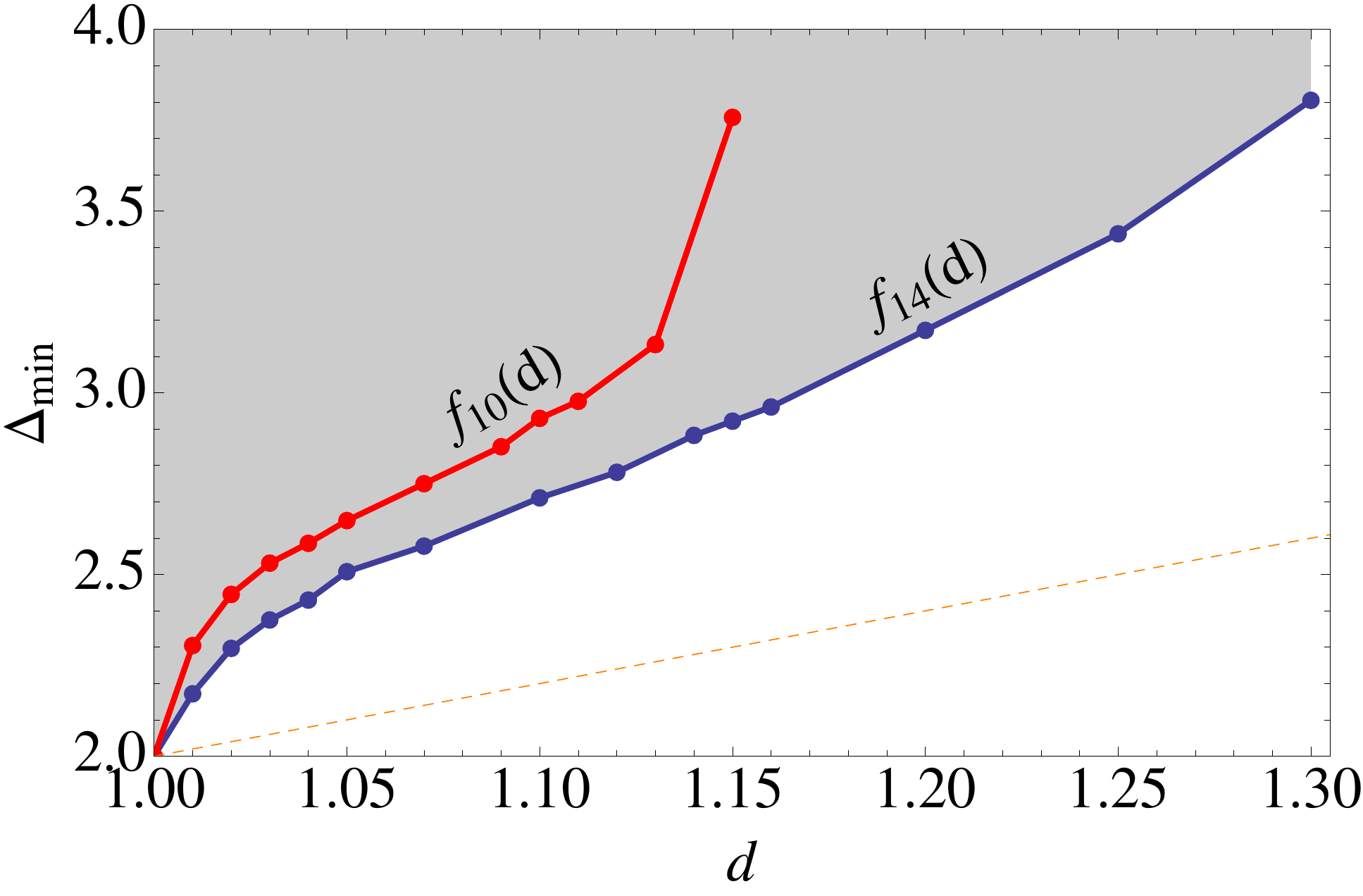}%
\includegraphics[scale=0.35]{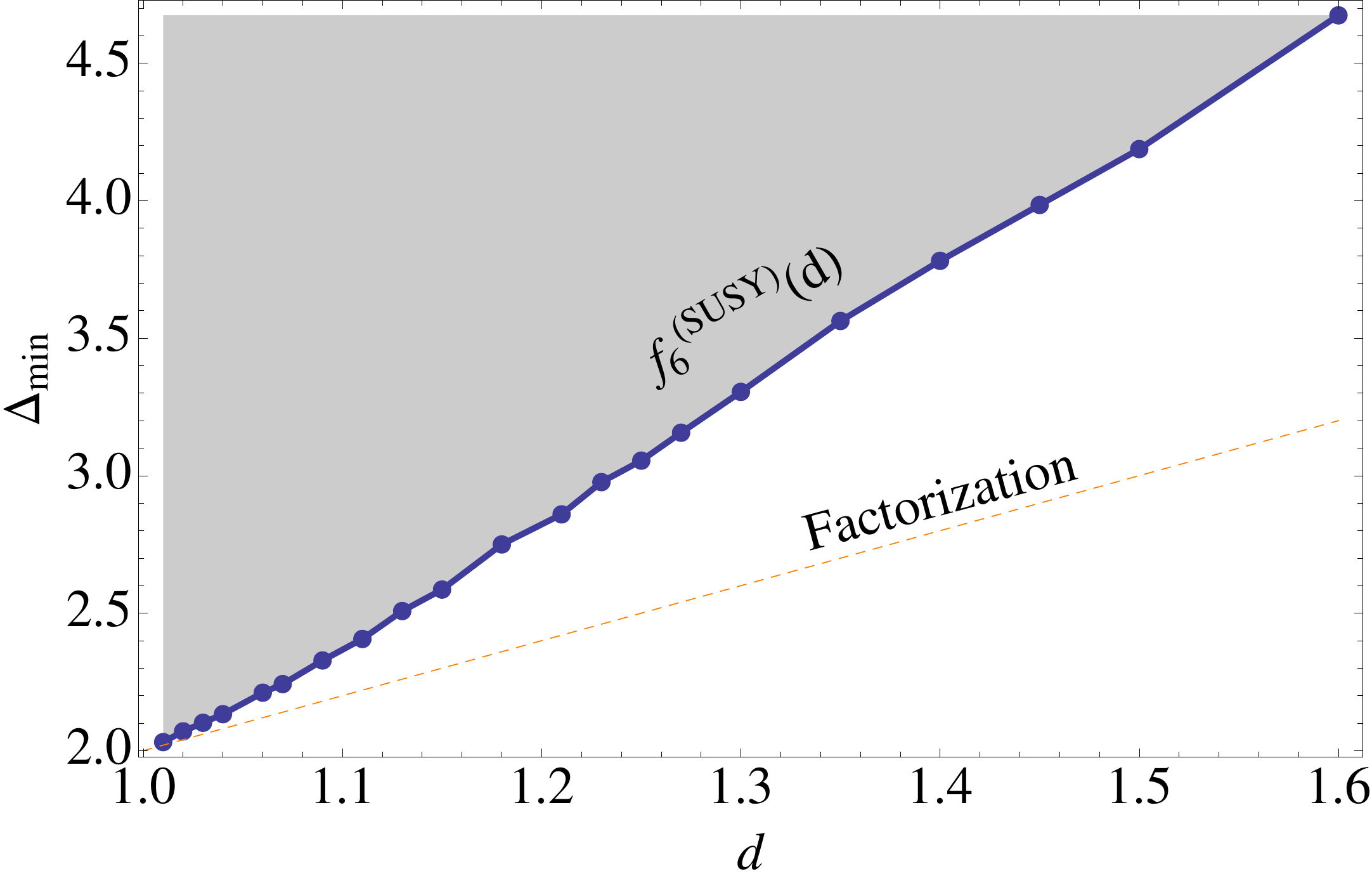}%
\caption[Bound on dimension of scalar non-chiral field in superconformal theories]{\footnotesize{\textit{Bound for the smallest dimension of a vector superfield appearing in the OPE of a chiral field with its conjugate. On the left: bound obtained obtained using only the first sum rule of (\ref{susyvecsumrule}); the bound $f_{10}(d)$ reproduces the results of \cite{poland}. On the right: bound obtained with 6 derivatives using the vectorial sum rule. Irregularities are due to the gap in dimension of the operators allowed by superconformal symmetry to appear in the $\Phi\times\Phi$ OPE.
}}}%
\label{susybound1}
\end{center}
\end{figure}
This time we denote the three structures as $V^{0\pm}$ , corresponding to operators with vanishing $R$-charge and even/odd spin, which appear in the $\phi\times\phi^\dagger$ OPE, and $V^2$, corresponding to operators  appearing in the $\phi\times\phi$ with $R$-charge twice the $\phi$ $R$-charge. The  fundamental difference among superconformal case and a pure $U(1)$ symmetric case is that supersymmetry relates some of the coefficient in the sum rule, grouping the sum in superconformal blocks. Moreover supersymmetry and the chirality of $\Phi$ restrict the allowed values of $\Delta$ in the sum. Recalling the definition of Sec. \ref{Superconformal blocks} we can write the vectorial sum rule in the schematic form:    
\be\label{susyvecsumrule}
\sum p_{\Delta,l}^{0+}
\underbrace{
\left(
\begin{array}
[c]{c}%
\mathcal F_{\Delta,l}\\
\widetilde{\mathcal F}_{\Delta,l}\\
\widetilde{\mathcal H}_{\Delta,l}%
\end{array}
\right)}_{\vec V^{0+}_{\Delta,l}}
 +\sum p_{\Delta,l}^{0-}
 \underbrace{
 \left(
\begin{array}
[c]{c}%
\mathcal F_{\Delta,l}\\
-\widetilde{\mathcal F}_{\Delta,l}\\
-\widetilde{\mathcal H}_{\Delta,l}%
\end{array}
\right)}_{\vec V^{0-}_{\Delta,l}}
  +\sum p_{\Delta,l}^{2}
 \underbrace{ 
  \left(
\begin{array}
[c]{c}%
0\\
F_{\Delta,l}\\
-H_{\Delta,l}%
\end{array}
\right) }_{\vec V^{2}_{\Delta,l}}
 =
\underbrace{
 \left(
\begin{array}
[c]{c}
1\\
1\\
-1
\end{array}
\right)}_{\vec V^{RHS}}
\ee %
In \cite{poland} a bound on the smallest dimension non-chiral scalar field with vanishing $R$-charge is derived making use of only the first sum rule of eq. (\ref{susyvecsumrule}). This method requires the use of an high number of derivatives, since no functional satisfying the suitable positivity properties exists for $N_{der}<10$. This is a consequence of the constraint incompleteness. Before exploiting the full power of the triple sum rule we review and improve the results of \cite{poland}. Hence we look for a linear functional defined on the functions $\mathcal F$ satisfying
  \ba
\Lambda\lbrack \mathcal F_{d,\Delta,l} ]\geq0\,,\,\,  &&  \text{for all }\Delta
\geq\Delta_{\min}\,(l=0)\nn\\
\phantom{-. }\text{and} \,\,  &&  \text{for all }\Delta
\geq l+2\,(l=1,2,3,4...)
\ea
The results are shown in Fig. \ref{susybound1} for different values of $N_{der}$. We observe that the bounds rapidly become weak as $d$ increases, ending in an absence of constraints on $\Delta_{min}$ for $d\gtrsim 1.2-1.3$. This behavior is understood as follows: as $d$ increases the convex cone spanned by the vectors of derivatives become wider and wider; at a certain value the projected cone fills the entire sub-space even without the need of $l=0$ vectors. In turns, the bounds on the scalar sector is absent. As we increase $N_{der}$ the value of $d$ where this degeneracy is reached grows. Instead using the entire set of constraints we do not find such a peculiarity. \\
Exploiting the triple sum rule requires again a linear functional defined on vector functions $\vec V^i_{\Delta,l}$. Unitarity bounds, superconformal symmetry and chirality restrict the positivity condition on $\Lambda$ (see Section. \ref{Representation of the Superconformal Algebra and unitarity bounds}, \ref{OPE of Chiral Superfields}, \ref{Superconformal blocks}):
 \ba
\Lambda\lbrack \vec V_{d,\Delta,l}^{0+}]\geq0\,,\,\,  &&  \text{for all }\Delta
\geq\Delta_{\min}\,(l=0)\\
\	\Lambda\lbrack \vec V_{d,\Delta,l}^{0\pm}]\geq0\text{ },\,\,  && \text{for all }\Delta\geq l+2\,(l=1,2,3,4,\ldots)\,.\nn\\
\	\Lambda\lbrack \vec V_{d,\Delta,l}^2]\geq0\,,\,   && \text{for all }\Delta= 2d +l\,(l=0,2,4\ldots)\,.\nn\\
\	\Lambda\lbrack \vec V_{d,\Delta,l}^2]\geq0\,,\,\,   && \text{for all }\Delta\geq |2d-3| +l+3\,(l=0,2,4\ldots)\,.\nn
\ea
The results obtained for $N_{der}=6$ are shown in the right plot of Fig. \ref{susybound1} . Compared to the plot on the left the bound has become more stringent, even exploiting a smaller number of derivatives. Moreover a fundamental difference is represented by the behavior of the bound close to the free theory. Compared to all the other cases so far investigated the bound approaches the free theory value linearly. This crucial difference could allow a direct comparison with perturbative calculations. Notice that the dimension of a chiral field is fixed by its  $R$-charge, which usually takes rational values. We didn't find any example in the literature where the dimension of $\Phi$ is sufficiently close to $1$ such that a perturbative correction to the dimension of the non-chiral operator $\Phi^\dagger\Phi$ has a chance to saturate the bound. In addition, most of the known $\mathcal N=1$ superconformal field theories have additional global symmetries (for instance the IR fixed points of \cite{Seiberg:1994pq}) under which the chiral superfield $\Phi$ is charged. Whenever this is the case the OPE $\phi\times\phi^\dagger$ contains  scalar field sitting in the same multiplet of a conserved global symmetry current. Since the current is conserved the dimension of the scalar field is constrained to be exactly 2.

\section{Central charge}
\label{Central charge}

So far we have used the sum rules to constrain
the maximal allowed gap in the scalar sector. In order to constrain the size
of the OPE coefficients $p_{\Delta,l}$, we proceed as in \cite{cr}.  For definiteness 
we consider a theory with a single sum rule and later on we will generalize.
Let us rewrite the sum rule extracting the contribution of one particular operator and transferring into the LHS: 
\begin{equation}
1-p_{\Delta_*,l_*}\,F_{d,\Delta_*,l_*}=\sum_{(\Delta,l)\neq(\Delta_*,l_*)}p_{\Delta
,l}F_{d,\Delta,l} \label{eq:sumrule1}%
\end{equation}
Assume we can find a linear functional such
that%
\begin{equation}
\Lambda\lbrack F_{d,\Delta,l}]\geq0 \label{eq:lambda2}%
\end{equation}
for all $\Delta,l$ and in addition $\Lambda[1]\geq 0$. Thus
\begin{equation}
\Lambda\lbrack1-p_{\Delta_*,l_*}\,F_{d,\Delta_*,l_*}]\geq 0. \label{eq:findt}%
\end{equation}
Since the functional is linear, Eq.~(\ref{eq:findt}) is satisfied for
\begin{equation}
p_{\Delta_*,l_*}\leq \Lambda\lbrack1]/\Lambda\lbrack F_{d,\Delta_*,l_*}],
\end{equation}
and for larger $p_{\Delta_*,l_*}$ the lhs of (\ref{eq:sumrule1}) becomes negative, while the rhs is positive. Thus we obtain the following result: each
functional $\Lambda$ satisfying (\ref{eq:lambda2}) gives a bound on the
maximal allowed value of $p_{\Delta_*,l_*}$:%
\begin{equation}
\max p_{\Delta_*,l_*}\leq\Lambda\lbrack1]/\Lambda\lbrack F_{d,\Delta_*,l_*}]\,.
\end{equation}
When there are more than a single sum rule the above condition is replaced by
\begin{equation}
\max \, p_{\Delta_*,l_*}\leq\frac{\Lambda\lbrack \vec V^{RHS}]}{\Lambda\lbrack \vec V_{d,\Delta_*,l_*}]}\,.
\label{eq:maxc42}%
\end{equation}
This bound can be optimized by choosing the functional judiciously. In particular, since the rhs of (\ref{eq:maxc42}) is homogeneous in the normalization of $\Lambda$ we can include in the positivity condition the constraint $ \Lambda[\vec V^{RHS}]=1 $
and maximize the quantity $\Lambda[\vec  V_{d,\Delta_*,l_*}]$ in order to get the most stringent bound\footnote{We are thankful to Slava Rychkov for suggesting an improved algorithm to address this issue.}.

Let us now concentrate on a particular OPE coefficient, the one associated to the 
energy momentum tensor. 
Before proceeding further it is useful to recall the relation between the central charge and OPE coefficients.
Being a conserved tensor of rank two, the energy-momentum tensor has dimension exactly $D$ in $D$ dimensions. Rescaling properly the energy momentum tensor to have it normalized according to \cite{Dolan:2003hv}  form  we can express the OPE coefficient $c_{D,2}$
 in terms of the central charge $C_{T}$ and
the dimension of $\phi$ \cite{Dolan:2003hv}:%
\begin{equation}
c_{D,2}=-\frac{Dd}{D-1}\frac{1}{\sqrt{C_{T}}}\,. \label{eq:c42}%
\end{equation}
The above equation is valid in arbitrary dimension $D$. In 2D the total central charge is defined as the sum of the central charges of holomorphic and antiholomorphic modes, $C_{T}=c+ \bar c $.
in the present conformal block convention  we have the relation:
\be\label{eq:p42}
\	p_{\Delta,l}=\frac{(c_{\Delta,l})^2§}{2^l}\,,\quad \Rightarrow \quad p_{D,2}=\frac{D^2d^2}{(D-1)^2}\frac{1}{4 C_T}
\ee
The above relation implies that for large $C_{T}$,
the contribution of the stress tensor to the 4-point function of $\phi$
decreases as $1/C_{T}$.

\subsection{Results for supersymmetric theories}
\label{Results for supersymmetric theories:cc}

\begin{figure}[h]
\begin{center}
\includegraphics[scale=0.35]{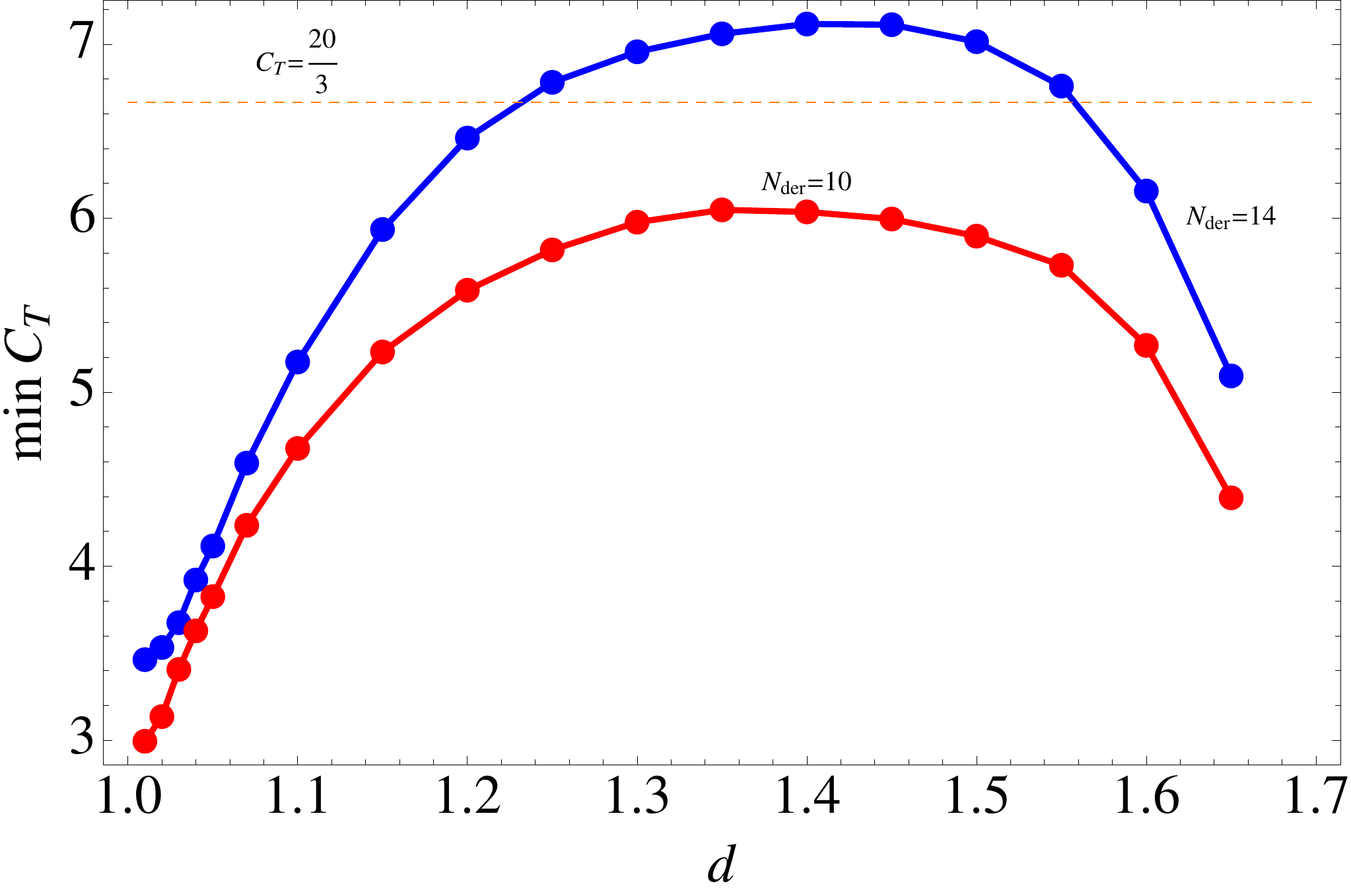} 
\includegraphics[scale=0.32]{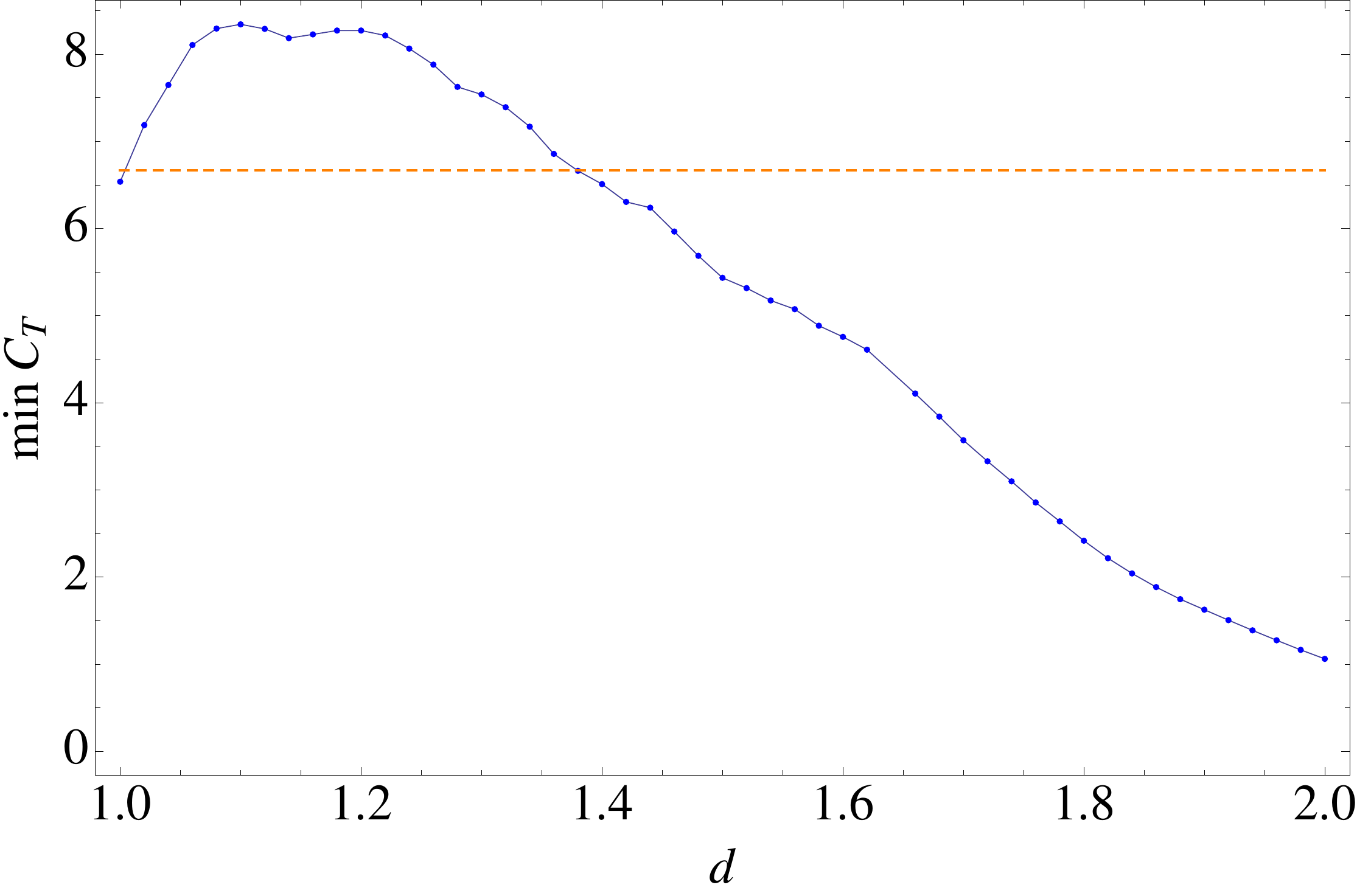}
\end{center}
\caption[Lower bound on the central charge in superconformal theories]{\footnotesize\textit{On the left: Lower bound on the central charge computed with $N_{der}=10$ (red) and $N_{der}=14$ (blue) using only one sum rule, the red bound reproduces the results of \cite{poland} . On the right: Lower bound on the central charge computed with $N_{der}=6$ using the vectorial sum rule. The dashed line corresponds to the central charge of a supersymmetric theory with one chiral superfield.}}\label{susycentralcharge}
\end{figure}

In superconformal field theories the energy momentum tensor is contained in the same super-multiplet of the $R$-current, which represents the lowest component. Hence the contribution of the energy-momentum tensor is encoded in the superconformal block  $\mathcal G_{3,1}(u,v)$. Using eq. (\ref{susypartialwave}), (\ref{susyCB}) and (\ref{eq:p42}) we can derive the relation between the conformal block coefficient and the central charge:
\be
\	p_{3,1}^{(susy)}=\frac{D^2d^2}{(D-1)^2}\frac{5}{4 C_T}\,.
\ee
Notice that using the decomposition of the four point function for a free complex scalar in therm of superconformal blocks (\ref{psusy}) we can verify that
\be
\	C_T=\frac{20}3\,,
\ee
which is the correct value for a free theory containing one complex scalar and one Weyl fermion \cite{Dolan:2003hv}. As for the bound on operator dimensions we can compute the lower bound on $C_T$ using only the information encoded in the first sum rule in (\ref{susyvecsumrule}) or, more correctly, using the whole set of equations. In the former case no bound can be extracted for $N_{der}<10$. The left plot in Fig. \ref{susycentralcharge} reports the result obtained with only one sum rule. The weaker bound reproduces the results of \cite{poland}, while the strongest one, derived with $N_{der}=14$, represents a numerical improvement. We see that using more computing effort we can constraint the central charge to be larger than the free theory value in a non negligible interval. This method however is not able to capture the correct behavior for $d\rightarrow1 $, limit in which the central charge is expected to reach the free value. 

The use of the entire set of crossing symmetry constraints allows instead to extract a bound even for small values of $N_{der}$.  Here we report the results obtained with $Nder=6$. Although the methods can be pushed to higher values of $N_{der}$ the bound becomes irregular and we decided not to present it. The right plot of Fig. \ref{susycentralcharge} shows the bound obtained imposing a positivity property of the form
 \ba\label{poscentralcharge}
\	\Lambda\lbrack \vec V_{d,\Delta,l}^{0\pm}]\geq0\text{ },\,\,  && \text{for all }\Delta\geq l+2\,(l=0,1,2,3,4,\ldots)\,.\nn\\
\	\Lambda\lbrack \vec V_{d,\Delta,l}^2]\geq0\,,\,   && \text{for all }\Delta= 2d +l\,(l=0,2,4\ldots)\,.\nn\\
\	\text{and}\,   && \text{for all }\Delta\geq |2d -3|+l+3\,(l=0,2,4\ldots)\,.\nn\\
\	\Lambda\lbrack \vec V^{RHS}]=1\,,\,
\ea	
where the the vectors $\vec V$ are defined in (\ref{susyvecsumrule}). As a first crucial difference with respect to the plot on the left we notice that the bound at $d=1$ is very close to the free values $20/3$  . In addition it remains above the free value for $d\leq 1.4$. For larger values of $d$ the bound decreases as usual \cite{r2, poland}. We expect however  the optimal bound to stay above the free value in all the open interval $1<d<2$.\\
A comparison with explicit models (\cite{poland}) shows that the bound is never saturated: for instance IR fixed points of SQCD theories in the conformal windows have a central charge at least a factor of 20 larger.  This discrepancy is somewhat expected since all these models contain additional global symmetries which increases the number of degrees of freedom and consequently the central charge. We expect that a combined use of vector sum rule for the proper global symmetry group and superconformal blocks could allow a non trivial comparison with these models.
\section{Conclusions}

In this work we produced numerical bounds investigating the crossing symmetry constraints derived in \cite{r4} for Conformal Field Theories with global symmetries. We restrict to two class of theories, both characterized by a set of three sum rules. As a first example we considered CFT's containing a scalar operator transforming in the fundamental representation of a global $SO(N)$. The operators entering the OPE of two such scalars can be organized in representation of the global symmetry group.
Restricting to $SO(4)$ we showed that independent bounds on the dimension of the first scalar singlet or scalar symmetric tensor  can be extracted. Moreover we investigated the dependence of the scalar singlet bound on $N$.

As a second example we consider the analysis of \cite{poland} for superconformal field theories containing at least one scalar chiral superfield with dimension $d$. The four point function of two chiral and two anti-chiral fields gives rise again to three crossing symmetry constraints. The novelty with respect to the original paper consists in the use of the full set of constraints. This allowed us to derive more severe bounds on the dimension of the first non-chiral scalar fields entering the OPE $\phi\times\phi^\dagger$. Moreover we extracted a lower bound on the central charge that lies above the value of the central charge of a free supersymmetric theory for a sizable range of values of $d$.

All the computations have been performed with the method described in details in \cite{r1,r2, poland} (and reviewed briefly in Section 4)  based on discretizing the set of constraints of the positivity property and reducing to a finite Linear Programming problem. Recently a new technique, based on a Semi-definite Programming Algorithm, has been developed \cite{SDPV}. The main conceptual difference resides in the fact that no discretization is required: the positivity on the continuous set of $\Delta$ can be directly imposed. Besides the confirmation and an improvement of the results presented here, the new techniques allows many additional investigations. For instance it can be shown than in CFT's with global $SO(N)$ symmetry the bound on the central charge scales with $N$, confirming the expectation that $C_T$ somehow measures the number of degrees of freedom. This analysis, extended to superconformal theories with $SU(N_f)\times SU(N_f)$ global symmetry could have interesting consequences: given the existence of exact results to compare with the numerical results can confirm analytical ones, disprove conjectures and guide further investigations on the  still obscure structure of CFT's.


\section*{Acknowledgments} 

I would like to thank Riccardo Rattazzi, David Simmons-Duffin and David Poland for interesting discussions and comments. I am particularly thankful to Slava Rychkov for the theoretical and technical discussions and for confirming some of the numerical results.

\end{document}